\pdfoutput=1

\documentclass[11pt]{article}

\usepackage{ACL2023}

\usepackage{times}
\usepackage{latexsym}

\usepackage[T1]{fontenc}

\usepackage[utf8]{inputenc}

\usepackage{microtype}

\usepackage{inconsolata}

\usepackage{graphicx} 
\usepackage{booktabs, array}
\usepackage{amsmath}
\usepackage{nicematrix}
\usepackage{multirow}
\usepackage{makecell}
\usepackage{pifont}
\usepackage{xcolor}
\usepackage{subcaption}
\usepackage{tcolorbox}
\usepackage{url}
\usepackage{graphicx}
\usepackage{enumitem}
\usepackage{placeins}
\usepackage{float}

\usepackage{amssymb}
\usepackage{dblfloatfix}
\usepackage{setspace}
\usepackage{colortbl}
\usepackage{adjustbox}
\usepackage{algorithm}
\usepackage{algpseudocode}
\usepackage{tabularx}
\usepackage{afterpage}
\usepackage{tcolorbox}
\tcbuselibrary{listings}

\definecolor{right}{RGB}{0,128,96}
\definecolor{wrong}{RGB}{192,0,32}
\definecolor{headerblue}{RGB}{218,232,252}
\definecolor{lightgray}{RGB}{245,245,245}

\newcommand{\Right}[1]{\textcolor{right}{#1}}
\newcommand{\Wrong}[1]{\textcolor{wrong}{#1}}
\newcommand{\cmark}{\Right{\ding{51}}}
\newcommand{\xmark}{\Wrong{\ding{55}}}

\newcommand{\todo}[1]{{\color{blue}TODO: #1}}

\title{\textsc{Multi$^3$IR}: A Benchmark for Multi-perspective Multi-domain Multi-modal Information Retrieval}

\author{
\quad \textbf{Seokwon Song}
\quad \textbf{Sohyeon Kim}
\quad \textbf{Gunhee Kim}\\
Seoul National University \\
\texttt{\small \{seokwon.song, sohyeon.kim\}@vision.snu.ac.kr}, 
\texttt{\small gunhee@snu.ac.kr} \\
}

\begin{document}
\maketitle
\begin{abstract}
Information retrieval (IR) increasingly targets open-ended queries that admit diverse perspectives. Existing IR benchmarks, however, focus primarily on closed-ended queries, while even open-ended benchmarks largely consist of queries whose supporting documents span a single subject domain and modality. We introduce \textsc{Multi$^3$IR}, a benchmark that evaluates how well retrievers cover the multifaceted perspectives of open-ended queries across diverse domains and modalities. It comprises 104.9K Stack Exchange queries, each annotated with perspective descriptions that capture the query's implicit viewpoints. We further propose \textsc{SPIN}, a parameter- and label-efficient method that learns noise vectors to steer embeddings toward diverse yet meaningful semantic directions. Experiments show that existing multimodal retrievers suffer from single-perspective bias, while \textsc{SPIN} substantially improves perspective coverage on \textsc{Multi$^3$IR} and generalizes well to unseen open-ended IR benchmarks. The dataset and experimental code are available at \url{https://github.com/seokwon99/Multi3IR}.
\end{abstract}

\section{Introduction}
\begin{figure}[t!]
\centering
\includegraphics[width=\linewidth]{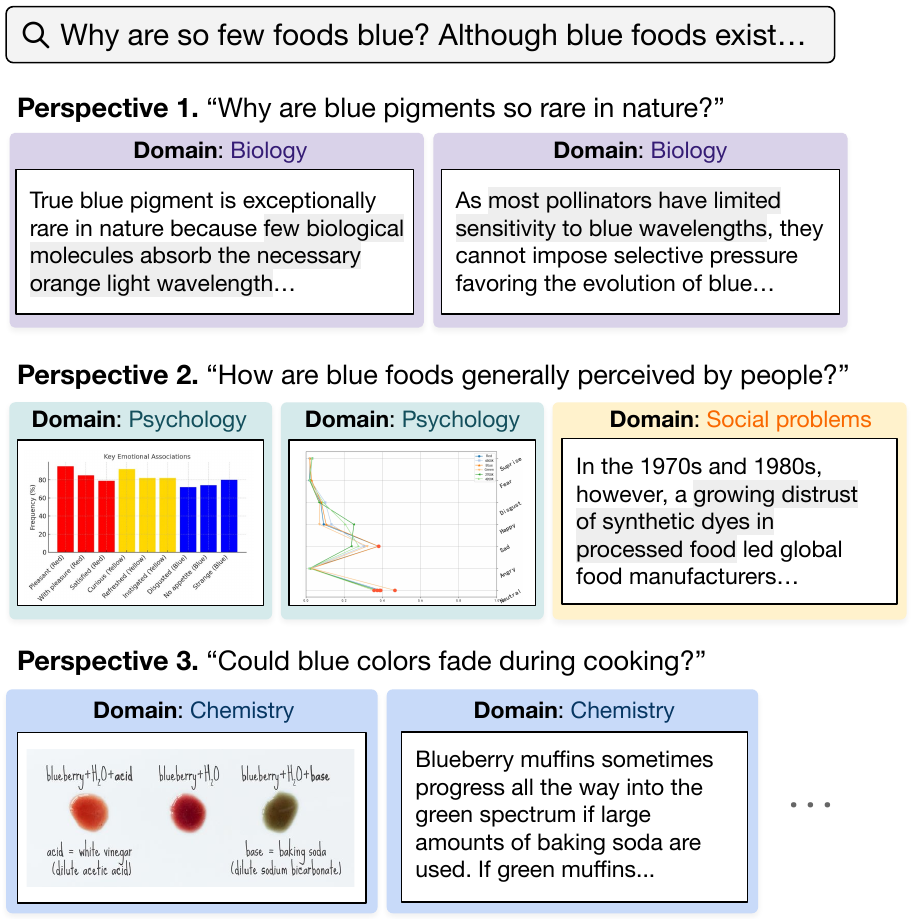}
\caption{A datapoint example of \textsc{Multi$^3$IR}. An open-ended query includes multiple implicit perspectives (three in this example), each paired with supporting documents from different domains and modalities.}
\label{fig:intro_figure}
\end{figure}

Information retrieval (IR) has expanded from factoid queries to open-ended queries that decompose into \textbf{perspectives}, complementary sub-queries that each capture a distinct facet. For example, as illustrated in Figure~\ref{fig:intro_figure}, the query ``Why are so few foods blue?'' implicitly raises perspectives such as ``Why are blue pigments so rare in nature?'' and ``How are blue foods generally perceived by people?'' These perspectives provide information from diverse subject domains (e.g., biology, psychology) and modalities (e.g., text, image), collectively contributing to a more comprehensive answer.

Recent efforts have advanced open-ended IR in several directions. On the evaluation side, some benchmarks~\cite{min2020ambigqa, zhao2024beyond, chen2025open} incorporate open-ended queries that admit multiple answers grounded in different documents. On the method side, retrieval diversification, including query expansion~\cite{gao2023precise, wang2023query2doc, zhang2024exploring} and multi-vector retrieval~\cite{khattab2020colbert, humeau2019poly}, seek to cover diverse aspects of a query rather than relying on a single dominant interpretation.

\newcommand{\tmark}{\todo{X.XX}}%
\newcommand{\citationformat}[1]{~\citeyearpar{#1}}
\begin{table}[t!]
    \centering
    \resizebox{\linewidth}{!}{%
    \begin{tabular}{lcccc}
        \toprule
         & \multirow{2}{*}{\makecell{\rule{0pt}{2.2ex}\#Q}}
         & \multicolumn{3}{c}{Diversity of each Query} \\
         \cmidrule(r){3-5}
         &  & \#SubQ & \#Domain & \#Modality \\
        \midrule
        \multicolumn{5}{l}{\textit{\textbf{Closed-ended IR}}} \\
        HotpotQA\citationformat{yang2018hotpotqa} & 113.0K & -- & 1.38 & 1.00 \\
        2WikiMHQA\citationformat{ho2020constructing} & 192.6K & 2.47 & 1.26 & 1.00 \\
        MultimodalQA\citationformat{talmor2021multimodalqa} & 29.9K & 1.83 & 1.20 & 1.23 \\
        FanOutQA\citationformat{zhu2024fanoutqa} & 1.0K & 6.92 & 1.67 & 1.00 \\
        WebQA\citationformat{chang2022webqa} & 41.7K & -- & 1.28 & 1.00 \\
        InfoSeek\citationformat{chen2023can} & 1.3M & -- & 1.00 & 1.00 \\
        M-BEIR\citationformat{wei2024uniir} & 1.2M & -- & 1.44 & 1.00 \\
        MRMR\citationformat{zhang2025mrmr} & 1.2K & -- & 1.03 & 1.05 \\
        \midrule
        \multicolumn{5}{l}{\textit{\textbf{Open-ended IR}}} \\
        AmbigQA\citationformat{min2020ambigqa} & 12.0K & 1.99 & 1.13 & 1.00 \\
        PIR\citationformat{zhao2024beyond} & 4.2K & 2.39 & 1.63 & 1.00 \\
        BeRDS\citationformat{chen2025open} & 3.2K & 2.21 & 1.30 & 1.00 \\
        \rowcolor{gray!10}
        Multi$^3$IR (Ours) & 104.9K & 4.98 & 3.34 & 1.91 \\
        \bottomrule
    \end{tabular}}
\caption{Comparison of IR benchmarks. \textit{Diversity per query} reports the average number of sub-queries (perspectives) per query, along with the number of distinct subject domains and modalities covered by their supporting documents. See Appendix~\ref{appendix:domain} for details on subject domain labeling.}
\label{tab:dataset_comparison}
\end{table}
However, current approaches remain limited in two key respects. First, as shown in Table~\ref{tab:dataset_comparison}, most IR benchmarks focus on closed-ended queries, overlooking information needs where quality depends on comprehensiveness or diversity. Even open-ended benchmarks largely consist of queries grounded in single-domain, text-only documents, leaving it unclear whether retrievers can provide comprehensive information across heterogeneous knowledge sources. Second, existing retrieval diversification methods incur substantial overhead: query expansion requires external LLMs at inference time, while multi-vector retrieval requires costly document-relevance labels for fine-tuning.

In this paper, we introduce \textsc{Multi$^3$IR}, an open-ended IR benchmark comprising 104.9K queries collected from Stack Exchange. For each query, we annotate a set of perspectives, each represented as a sub-query paired with the documents that support it. As shown in Table~\ref{tab:dataset_comparison}, each query in our benchmark requires comprehensive information spanning 3.34 domains and 1.91 modalities on average. We further propose \textsc{SPIN}, a parameter- and label-efficient IR training method that learns noise vectors to diversify the output of a frozen retriever into multiple perspective-aware embeddings, without requiring costly document-level annotations.

Our main contributions are as follows:
\begin{enumerate}
    \item We propose \textsc{Multi$^3$IR}, a large-scale benchmark of 104.9K open-ended queries that require comprehensive information across multiple subject domains and modalities.
    \item We identify \textit{single-perspective bias} in current multimodal retrievers, where retrieval for an open-ended query concentrates on a few dominant perspectives while neglecting the rest.
    \item We propose \textsc{SPIN}, a parameter- and label-efficient training method that achieves higher perspective coverage than existing diversification methods.
\end{enumerate}

\begin{figure*}[!t]
    \centering
    \includegraphics[width=0.95\linewidth]{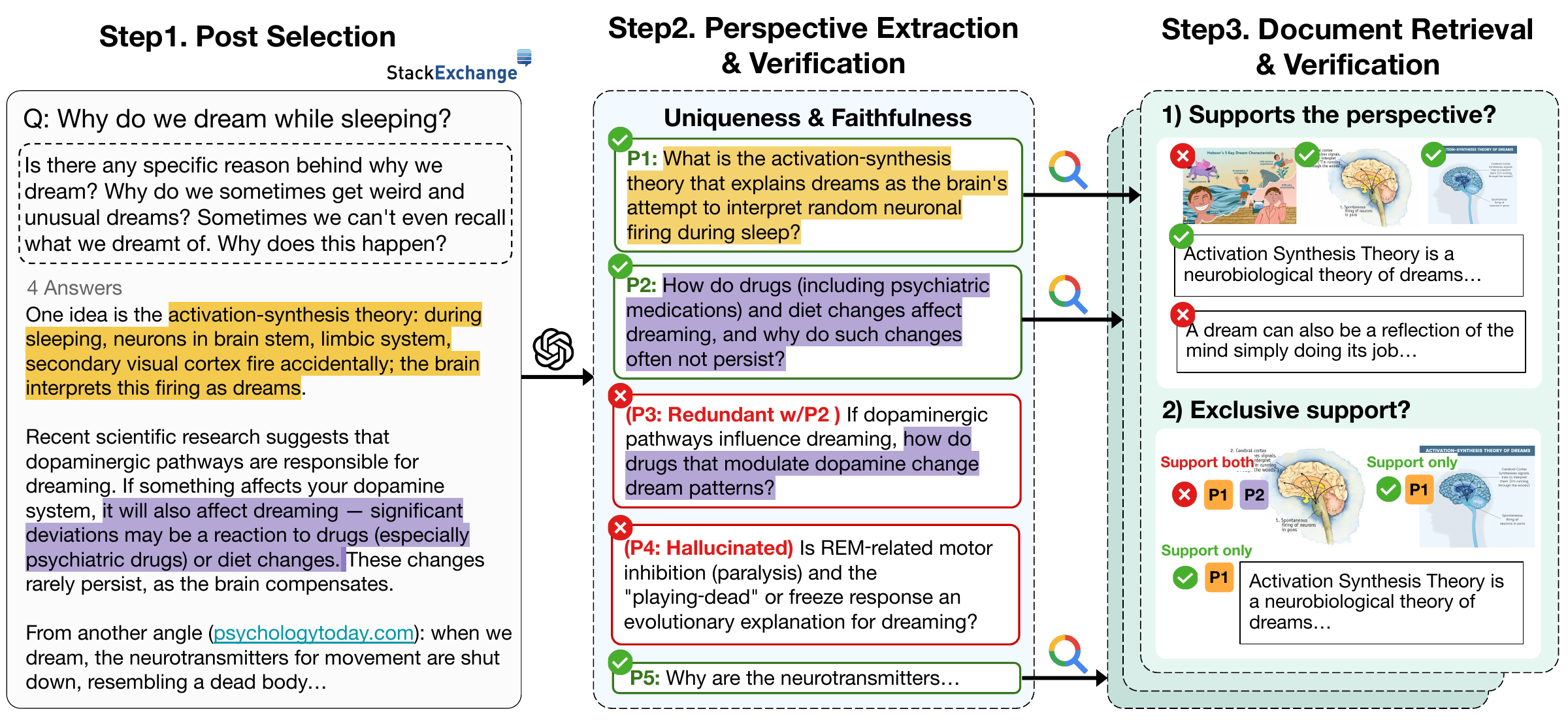}
    \caption{Overview of our data construction pipeline (\S~\ref{data_construction}).} \label{fig:data_construction}
\end{figure*}

\section{Related Work}
\subsection{Open-ended Information Retrieval}
Open-ended IR benchmarks examine whether a retriever can gather documents covering diverse information needs within a question. Early work such as AmbigQA~\cite{min2020ambigqa} introduced questions with multiple valid interpretations, with each disambiguation grounded in Wikipedia documents, but evaluates question answering rather than retrieval. PIR~\cite{zhao2024beyond} extends this setting to retrieval, evaluating whether a retriever can return documents relevant to an explicitly specified perspective within a query. Since users rarely state their perspectives, BeRDS~\cite{chen2025open} instead gives the retriever only the original question and evaluates whether the retrieved documents cover all annotated perspectives. However, these benchmarks largely consist of queries grounded in single-domain, text-only documents, limiting their ability to evaluate comprehensive retrieval across diverse knowledge sources. \textsc{Multi$^3$IR} addresses this gap by collecting supporting documents across multiple domains in both text and image modalities.

\subsection{Retrieval Diversification}
Dense retrievers rank documents by similarity to a single query vector, which cannot be close to all relevant documents when they are scattered across the embedding space~\cite{chen2025beyond, weller2025theoretical}. Prior work introduces diversity at three stages. \textit{Post-retrieval} re-ranks an initial candidate pool to promote diverse results~\cite{carbonell1998use, kulesza2012determinantal}; since they only reorder the first-stage pool, documents missing from it remain unretrievable. \textit{Query expansion} rewrites the query before retrieval, generating hypothetical documents~\cite{gao2023precise} or pseudo-answers~\cite{wang2023query2doc,zhang2024exploring}, at the cost of an LLM call for every query. \textit{Multi-vector retrieval} represents a query with multiple embeddings~\cite{khattab2020colbert,humeau2019poly}, expanding the regions it can cover, but training them relies on costly document-level relevance annotations, especially for open-ended queries spanning multiple perspectives. \textsc{SPIN} addresses this by training a multi-vector retriever using only perspective descriptions, improving perspective coverage without document-level annotations or query-time LLM inference.

\section{The \textsc{Multi$^3$IR} Benchmark}

\subsection{Task Formulation}
Open-ended query $Q$ involves multiple implicit perspectives (Figure~\ref{fig:intro_figure}). We define a set of perspectives $P$, where each $p_i \in P$ is a textual description grounded in documents $D_i$. Given $Q$ and a corpus $C$, a retriever $f(\cdot)$ that maps inputs to a representation space returns the top-$k$ documents as

\begin{equation}
D^{*}(k) = \mathop{\mathrm{Top\text{-}}k}_{d \in C} f(Q)^{\top} f(d),
\end{equation}

The retrieval objective is to cover every perspective $p_i \in P$ by retrieving at least one document from its corresponding supporting set $D_i$.

\subsection{Dataset Construction}
\label{data_construction}

\begin{figure}[!t]
\centering
\begin{subfigure}[t]{\linewidth}
\centering
\resizebox{\linewidth}{!}{%
\begin{tabular}{lc}
\toprule
\textbf{Statistic} & \textbf{Number} \\
\midrule
\textbf{Query} &  \\
\; - Total & 104,916 \\
\; - Category type & 5 \\
\; - Avg. tokens & 155.8 \\
\midrule
\textbf{Perspective} &  \\
\; - Total & 521,739 \\
\; - Avg. tokens & 27.1 \\
\midrule
\textbf{Supporting Document} &  \\
\; - Total & 1,012,061 \\
\; - Modality type & 2 \\
\; - Domain type & 94 (of 100) \\
\midrule
\textcolor{gray}{Category of Queries} & \\
\; - Culture \& Recreation & 34,776 (33.5\%) \\
\; - Science & 20,344 (19.6\%) \\
\; - Business & 19,132 (18.4\%) \\
\; - Technology & 15,756 (15.2\%) \\
\; - Life \& Arts & 13,869 (13.4\%) \\
\midrule
\textcolor{gray}{Domain of Documents} & \\
\; - Computer science \& systems & 197,451 (19.5\%) \\
\; - Management \& public relations & 94,731 (9.4\%) \\
\; - Law & 69,191 (6.8\%) \\
\; - Engineering & 67,122 (6.6\%) \\
\; - Others (90 domains) & 583,449 (57.7\%) \\
\midrule
\textcolor{gray}{Modality of Documents} & \\
\; - Text & 631,377 (62.4\%) \\
\; - Image & 380,684 (37.6\%) \\
\bottomrule
\end{tabular}}
\caption{Overall dataset statistics. Gray sections report ratios sorted in descending order.}
\label{tab:dataset_stats}
\end{subfigure}

\begin{subfigure}[t]{\linewidth}
\centering
\includegraphics[width=\linewidth]{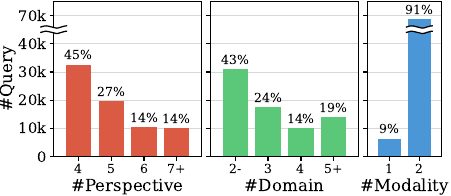}
\caption{Train query distribution by the number of perspectives, document domains, and document modalities.}
\label{fig:query_distributions}
\end{subfigure}
\caption{Overall statistics and distribution analysis.}
\label{fig:dataset_stats_all}
\end{figure}

As illustrated in Figure~\ref{fig:data_construction}, we construct the dataset using an automated three-stage pipeline. After that, we conduct human annotation to construct test set. Overall dataset statistics are presented in Figure~\ref{fig:dataset_stats_all}.

\paragraph{Selecting posts.}
We collect posts from Stack Exchange, a rich source of multi-answer threads with diverse perspectives. We use 77 sites across five categories, retain posts with more than 3 answers, and sample uniformly across categories. For each post, $Q$ is the concatenated title and body, and $A$ is the set of answers. See Appendix~\ref{appendix:raw_data_collection} for details.

\paragraph{Perspective extraction and verification.}
We extract perspectives $\hat{P} = \{\hat{p}_{1}, \hat{p}_{2}, \ldots\}$ from $A$ using GPT-5-mini, then verify them as follows:

\begin{enumerate}
    \item \textbf{Uniqueness.} We encode all perspectives with all-MiniLM-L6-v2~\cite{reimers2019sentence} and discard any $\hat{p}_{j}$ whose maximum cosine similarity to $\hat{P} \setminus \{\hat{p}_{j}\}$ exceeds $0.8$.
    \item \textbf{Faithfulness.} We use Bespoke-MiniCheck-7B~\cite{tang-etal-2024-minicheck} to produce a binary entailment judgment between each $\hat{p}_j$ and $A$, and discard any $\hat{p}_j$ judged as not entailed.
\end{enumerate}

We remove questions with fewer than four perspectives, yielding 104.9K questions and 521.7K perspectives. See Appendix~\ref{appendix:perspective_extraction_and_filtering} for details.

\paragraph{Document retrieval and verification.}
We collect documents supporting each perspective from two sources: Google Image Search for images and the Colossal Clean Crawled Corpus (C4)~\citep{raffel2020exploring} for text. Using each $\hat{p}_{i}$ as a query, we retrieve the top-10 documents from each and define $\hat{D}_{i}$ as their union. Since retrieved documents may support multiple perspectives, we enforce \textit{exclusive support}: using Qwen3-VL-30B-A3B-Instruct~\citep{qwen3technicalreport}, we verify each $\hat{d} \in \hat{D}_{i}$ against $\hat{p}_{i}$ and the other perspectives, and define $D_{i}$ as those supporting $\hat{p}_{i}$ alone. Finally, we label each document in $D_{i}$ with a subject domain using EAI-Distill-0.5b~\citep{ai2025essentialwebv1024ttokens}, yielding 1.01M multimodal documents with 3.34 domains and 1.91 modalities per query on average. See Appendix~\ref{appendix:document_retrieval_and_filtering} for retrieval and verification details, and Appendix~\ref{appendix:domain} for domain labeling.

\subsection{Human Verification}

We verify each sample in two stages. Annotators first judge (Q1a) whether the perspective is relevant to the question, (Q1b) which other perspective is closest to it, and (Q1c) whether the two are semantically redundant. They then rate whether the document fully, partially, or does not support (Q2a) the target perspective and (Q2b) the closest perspective from Q1b. Nearly all perspectives are relevant (99.1\%) and unique (96.2\%), with 90.2\% of documents exclusively supporting their target perspective. We retain perspectives passing Q1 with their supporting documents, yielding a test split of 1.0K queries, 4.8K perspectives, and 12.5K supporting documents. See Appendix~\ref{appendix:human_annotation} for details.


\section{Approach}
\begin{figure}[t!]
    \centering
    \includegraphics[width=\linewidth]{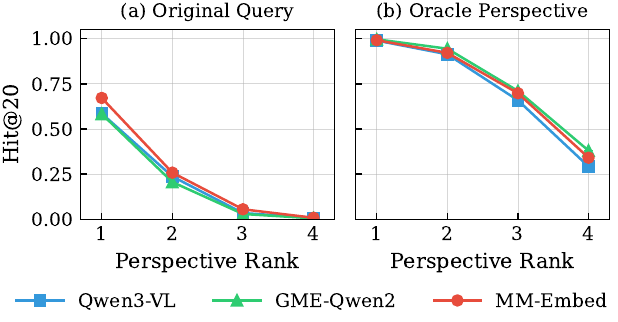}
    \caption{Preliminary analysis. Hit@20 is measured for each perspective and perspectives are ranked accordingly. (a) Naive retrieval using the original query. (b) Multi-query retrieval using ground-truth perspective descriptions, with results aggregated via Round Robin.}
    \label{fig:motivation_exp}
\end{figure}
\begin{algorithm*}[t]
\caption{\textsc{SPIN} Optimization}\label{alg:spin-opt}
\small
\textbf{Require:} Frozen encoder $M$ with $L$ layers, training set $\mathcal{X}$, learnable noise vectors $\mathcal{E} = \{\boldsymbol{\varepsilon}_1, \ldots, \boldsymbol{\varepsilon}_m\}$, injection layer $l$, sigmoid temperature $\tau$, bias $b$
\begin{algorithmic}[1]
\For{batch $\mathcal{B} = \{(q_n, \mathcal{P}_n)\}_{n=1}^{N} \subset \mathcal{X}$} \Comment{$q_n$: $n$-th query, $\mathcal{P}_n$: corresponding perspective descriptions}
    \For{$n = 1, \ldots, N$}
        \State $\mathbf{h}_l^n \leftarrow M_{0 \rightarrow l}(q_n)$ \Comment{$\mathbf{h}_l^n$: internal representation of $q_n$ at layer $l$}
        \State $\mathbf{S}_n \leftarrow \{M_{l \rightarrow L}(\mathbf{h}_l^n + \boldsymbol{\varepsilon}) \mid \boldsymbol{\varepsilon} \in \mathcal{E}\}$ \Comment{$\mathbf{S}_n$: steered embeddings for $q_n$}
        \State $\mathbf{T}_n \leftarrow \{M(p) \mid p \in \mathcal{P}_n\}$ \Comment{$\mathbf{T}_n$: target embeddings for $q_n$}
    \EndFor
    \State $p_{\mathbf{s}}(\mathbf{t}) \leftarrow \sigma\!\left((\cos(\mathbf{s}, \mathbf{t}) - b) / \tau\right)$ \Comment{$p_{\mathbf{s}}(\mathbf{t})$: probability that steered $\mathbf{s}$ covers target $\mathbf{t}$}
    \vspace{2pt}
    \State $\mathcal{L}_{\text{pos}} \leftarrow -\frac{1}{N} \sum_{n=1}^{N} \frac{1}{|\mathbf{T}_n|}\sum_{\mathbf{t} \in \mathbf{T}_n} \log\!\left(1 - \prod_{\mathbf{s} \in \mathbf{S}_n} \bigl(1 - p_{\mathbf{s}}(\mathbf{t})\bigr)\right)$
    \State $\mathcal{L}_{\text{neg}} \leftarrow -\frac{1}{N} \sum_{n=1}^{N} \frac{1}{|\mathbf{T}_{\neq n}|}\sum_{\mathbf{t} \in \mathbf{T}_{\neq n}} \sum_{\mathbf{s} \in \mathbf{S}_n} \log\!\bigl(1 - p_{\mathbf{s}}(\mathbf{t})\bigr)$ \Comment{$\mathbf{T}_{\neq n}$: in-batch negative targets}
    \State Minimize $\mathcal{L} = \mathcal{L}_{\text{pos}} + \mathcal{L}_{\text{neg}}$
\EndFor
\end{algorithmic}
\end{algorithm*}

\subsection{Preliminary Analysis}
Open-ended queries often admit multiple perspectives, yet whether existing retrievers adequately capture such diversity remains unclear. We investigate this on our test set using three multimodal retrievers: MM-Embed-8B~\citep{lin2024mm}, GME-Qwen2-VL-7B-Instruct~\citep{zhang2024gme}, and Qwen3-VL-Embedding-8B~\citep{qwen3vlembedding}.

\paragraph{Do retrievers cover multiple perspectives?}
We first examine perspective coverage (Figure~\ref{fig:motivation_exp}(a)). Given the original query, we retrieve the top 20 documents and check whether at least one supporting document is retrieved for each perspective. We find that retrieval is heavily concentrated on a single dominant perspective, while the remaining ones receive substantially lower scores, revealing a \textit{single-perspective bias}.

\paragraph{Where does the bottleneck originate?}
We next ask whether the bias stems from the query embedding failing to capture other perspectives or from poor alignment between the query and the document embeddings of their supporting documents (Figure~\ref{fig:motivation_exp}(b)). When we instead use perspective descriptions as queries and aggregate the results via Round Robin, the previously missed documents become retrievable, yielding substantially more balanced performance across perspective ranks. This indicates that the bottleneck lies in the query embedding rather than the document space, motivating perspective-guided learning with multiple query vectors aligned to explicit perspective descriptions.




\subsection{Perspective-Guided Learning}
\label{perspective_guided}

To enable the retriever to interpret queries in diverse ways, we propose \textsc{SPIN} (\underline{S}teering \underline{P}erspectives by \underline{I}njecting \underline{N}oise), which requires only perspective descriptions and no document relevance annotations. Prior work~\citep{skean2024does, skean2025layer} shows that intermediate transformer layers preserve richer semantic information, whereas the final layers converge toward a single fixed interpretation. Motivated by this, \textsc{SPIN} injects learnable noise vectors at an intermediate layer of a frozen retriever to steer its embeddings toward diverse perspectives.

\paragraph{Optimization process.} 
We describe the optimization procedure in Algorithm~\ref{alg:spin-opt}. For each query $q_n$, we inject each learnable noise vector $\boldsymbol{\varepsilon} \in \mathcal{E}$ into the hidden representation of $q_n$ at layer $l$ and forward it through the remaining layers, yielding the steered embeddings $\mathbf{S}_n$. We then encode the perspective descriptions $\mathcal{P}_n$ with the same frozen encoder $M$ into target embeddings $\mathbf{T}_n$, and optimize $\mathcal{E}$ so that $\mathbf{S}_n$ aligns with $\mathbf{T}_n$.

Since both $\mathbf{S}_n$ and $\mathbf{T}_n$ consist of multiple vectors, aligning the two sets requires a many-to-many optimization. Rather than imposing a fixed one-to-one assignment (e.g., Hungarian matching~\citep{kuhn1955hungarian}), we formulate alignment as a probabilistic coverage problem. Specifically, each $\mathbf{s} \in \mathbf{S}_n$ independently produces a coverage probability for a target $\mathbf{t} \in \mathbf{T}_n$, and these probabilities are combined via a noisy-OR, so that $\mathbf{t}$ is considered covered if at least one $\mathbf{s}$ aligns with it. The positive loss $\mathcal{L}_{\text{pos}}$ encourages every $\mathbf{t} \in \mathbf{T}_n$ to be covered by \emph{any} $\mathbf{s} \in \mathbf{S}_n$, whereas the negative loss $\mathcal{L}_{\text{neg}}$ requires \emph{all} $\mathbf{s} \in \mathbf{S}_n$ to reject in-batch negatives from other queries, $\mathbf{T}_{\neq n}$.

\paragraph{Inference.} After optimization, given a query $q$, we obtain $m$ embeddings by injecting each optimized noise vector at layer $l$:
\begin{equation}
    \mathbf{E}^q = \{R_{l \rightarrow L}\big(R_{0 \rightarrow l}(q) + \varepsilon) \mid \varepsilon \in \mathcal{E}\}.
\end{equation}
For each embedding $\mathbf{e} \in \mathbf{E}^q$, we independently retrieve a ranked list $\mathcal{R}_\mathbf{e}$ from the document corpus $\mathcal{C}$ via Maximum Inner Product Search:
\begin{equation}
    \mathcal{R}_\mathbf{e} = \text{argsort}_{d \in \mathcal{C}} \; \mathbf{e} \cdot M(d).
\end{equation}
Then, the $m$ ranked lists are aggregated via Round Robin:
\begin{equation}
    \mathcal{R}^q = \text{RoundRobin}(\mathcal{R}_{\mathbf{e}_1}, \mathcal{R}_{\mathbf{e}_2}, \dots, \mathcal{R}_{\mathbf{e}_m}).
\end{equation}
At each round $r = 1, 2, \dots$, we iterate through $(\mathcal{R}_{\mathbf{e}_1}, \dots, \mathcal{R}_{\mathbf{e}_m})$ in order and append the rank-$r$ document of each list to $\mathcal{R}^q$, skipping duplicates. The procedure terminates once $|\mathcal{R}^q| = k$.

\section{Experiments}
\subsection{Evaluation Metrics}

We evaluate perspective coverage using two metrics. Hard coverage checks whether the retrieved documents include any annotated supporting document for each perspective, offering reproducibility and scalability. Since our annotations may miss relevant documents, soft coverage uses GPT-5-mini to assess whether each retrieved document supports a given perspective, recovering evidence that hard coverage would treat as a miss. These hard and soft coverage metrics are defined by
\begin{equation}
\mathrm{HC@k} = \frac{1}{|P|} \sum_{i=1}^{|P|} \mathbf{1}\!\left[D^{*}(k) \cap D_{i} \neq \emptyset\right],
\end{equation}
\begin{equation}
\mathrm{SC@k} = \frac{1}{|P|} \sum_{i=1}^{|P|} \max_{d \in D^{*}(k)} \mathrm{Judge}(p_i, d),
\end{equation}
where $\mathrm{Judge}(p_i, d) \in \{0, 1\}$ indicates whether $p_i$ is fully supported by $d$. Specifically, we report HC at $k \in \{5, 10, 20, 100\}$ and SC at $k \in \{5, 10, 15, 20\}$, with SC limited to smaller $k$ due to the API cost. See Appendix~\ref{appendix:evaluation} for evaluation details, including human agreement with the GPT-5-mini judge.

\begin{table*}[t!]
    \centering
    \resizebox{\linewidth}{!}{%
    \begin{NiceTabular}{lccccccccccc}[colortbl-like]
        \toprule
         & & & & \multicolumn{4}{c}{\textbf{Hard Coverage (HC)}} & \multicolumn{4}{c}{\textbf{Soft Coverage (SC)}} \\
         \cmidrule(r){5-8}\cmidrule(r){9-12}
        \textbf{Method} & $\mathbf{m}$ & \textbf{FT-Doc} & \textbf{FT-Persp} & @5 & @10 & @20 & @100 & @5 & @10 & @15 & @20 \\
            \midrule
        
        \multicolumn{12}{l}{\textit{\textbf{MM-Embed}}} \\
        \text{Naive} & 1 & \xmark & \xmark & 21.07 & 28.83 & 35.50 & 52.26 &  29.43 & 39.52 & 45.85 & 49.53 \\
        \text{Naive} & 1 & \cmark & \xmark & 20.95 & 28.38 & 35.94 & 53.96 & 31.60 & 41.55 & 47.81 & 51.80 \\
        \text{LLM-Expansion} & 5 & \xmark & \cmark & 15.63 & 22.11 & 30.11 & 47.37 & 27.16 & 37.13 & 43.25 & 47.32 \\
        \text{SPIN} & 5 & \xmark & \cmark & \textbf{25.75} & \textbf{35.24} & \textbf{41.24} & \textbf{59.17} & \textbf{49.22} & \textbf{59.05} & \textbf{64.31} & \textbf{67.54} \\
        \rowcolor{gray!20}\text{Oracle} & 4.6 & \xmark & \xmark & 34.00 & 44.41 & 53.79 & 71.71 & 57.55 & 70.80 & 76.31 & 79.29 \\
        \midrule
        \multicolumn{12}{l}{\textit{\textbf{GME-Qwen2}}} \\
        \text{Naive} & 1 & \xmark & \xmark & 16.18 & 21.81 & 27.92 & 42.08 & 26.59 & 35.40 & 40.58 & 44.86 \\
        \text{Naive} & 1 & \cmark & \xmark & 17.69 & 24.20 & 31.68 & 50.84 & 26.28 & 36.19 & 41.66 & 46.38 \\
        \text{LLM-Expansion} & 5 & \xmark & \cmark & 18.96 & 25.84 & 33.60 & 51.04 & 27.07 & 38.01 & 44.13 & 47.99 \\
        \text{ARE} & 5 & \cmark & \xmark & 22.22 & 30.40 & 39.23 & \textbf{59.35} & 30.69 & 41.53 & 48.72 & 53.55 \\
        \text{SPIN} & 5 & \xmark & \cmark & \textbf{24.95} & \textbf{34.02} & \textbf{41.39} & 58.62 & \textbf{45.64} & \textbf{56.00} & \textbf{61.20} & \textbf{64.81} \\
        \rowcolor{gray!20}\text{Oracle} & 4.6 & \xmark & \xmark & 45.44 & 57.12 & 65.07 & 77.68 & 59.06 & 72.12 & 77.46 & 79.95 \\
        \midrule
        \multicolumn{12}{l}{\textit{\textbf{Qwen3-VL}}} \\
        \text{Naive} & 1 & \xmark & \xmark & 18.43 & 25.55 & 34.16 & 48.28 & 26.75 & 36.22 & 41.74 & 45.63 \\
        \text{Naive} & 1 & \cmark & \xmark & 20.71 & 28.06 & 37.47 & 59.37 & 30.76 & 41.41 & 47.52 & 52.44 \\
        \text{LLM-Expansion} & 5 & \xmark & \cmark & 18.44 & 25.33 & 33.11 & 50.95 & 27.06 & 37.75 & 43.80 & 48.31 \\
        \text{ARE} & 5 & \cmark & \xmark & 24.75 & 33.36 & 42.71 & 61.60 & 33.50 & 44.73 & 51.37 & 56.01 \\
        \text{SPIN} & 5 & \xmark & \cmark & \textbf{26.44} & \textbf{37.83} & \textbf{45.55} & \textbf{63.44} & \textbf{49.62} & \textbf{59.35} & \textbf{64.47} & \textbf{67.47} \\ 
        \rowcolor{gray!20}\text{Oracle} & 4.6 & \xmark & \xmark & 42.96 & 52.31 & 60.83 & 73.76 & 58.18 & 70.04 & 75.31 & 78.14 \\
        \bottomrule
    \end{NiceTabular}} 
    \caption{Results of perspective coverage (\%) on the test split of \text{Multi$^3$IR-Bench} (1.01M candidates). \textit{m} denotes the number of query embeddings. \textit{FT-Doc} denotes fine-tuning with positive documents per query, and \textit{FT-Persp} denotes fine-tuning with perspective descriptions per query. \textit{Oracle} in gray background represents upper-bound performance for each retriever and is not included in the rankings.}
    \label{tab:main_results}
\end{table*}

\subsection{Baselines}

We experiment with several retrieval diversification methods on three state-of-the-art multimodal retrievers: GME-Qwen2-VL-7B-Instruct~\cite{zhang2024gme}, MM-Embed-8B~\cite{lin2024mm}, and Qwen3-VL-Embedding-8B~\cite{qwen3vlembedding}. For multi-vector retrieval methods ($m > 1$), results are aggregated via round robin. We use 72K instances as the training set and 32K as the validation set, and document embeddings remain frozen throughout. See Appendix~\ref{appendix:implementation} for implementation details.

\paragraph{Naive.} We evaluate the retrievers in both zero-shot and fine-tuned ways. For fine-tuning, we concatenate positive documents across all perspectives per query into a single positive set, and train the retriever with in-batch negatives using InfoNCE~\cite{oord2018representation}.

\paragraph{LLM-Expansion.} We diversify the query space by prompting an LLM with the instruction ``Generate \{m\} search queries related to: \{user\_query\}'' to obtain \(m\) distinct query reformulations. To adapt the model to our setting, we fine-tune Qwen3-4B~\cite{qwen3technicalreport} to generate the annotated perspective descriptions corresponding to each query, where $m$ is set to the number of annotated perspectives during training. At inference time, we fix $m = 5$ for consistency across queries.

\paragraph{Auto-Regressive Embedding (ARE).} We implement the ARE method~\cite{chen2025beyond, huo2026causalembed} in our setting, which generates multi-vector embeddings auto-regressively. During training, the retriever takes the query tokens followed by supporting document embeddings and produces one query embedding per position. The query embeddings are aligned with gold documents via Hungarian matching, and InfoNCE is applied with in-batch negatives. At inference, we fix $m=5$ and perform $m$ sequential forward passes. ARE cannot be applied to the encoder-based retriever MM-Embed, so we exclude it from the experiments.

\paragraph{SPIN.} We inject $m$ learnable noise vectors at an intermediate layer to steer final-layer embeddings, requiring only perspective descriptions without document relevance annotations. To account for different backbone depths, we set the injection point at $l=0.5L$, corresponding to layers $16$, $14$, and $18$ for MM-Embed ($L=32$), GME-Qwen2 ($L=28$), and Qwen3-VL ($L=36$), respectively. See \S~\ref{perspective_guided} for details.

\paragraph{Oracle.} We use oracle perspective descriptions as retrieval queries, serving as an upper bound for perspective-guided retrieval.

\subsection{Results on the \textsc{Multi$^3$IR}}
\label{main_results}
We report the main results on \textsc{Multi$^3$IR} in Table~\ref{tab:main_results}, 
where all baselines retrieve candidates from the full 1.01M gold document pool. We adopt 
HardCoverage@$k$ (HC@$k$) and SoftCoverage@$k$ (SC@$k$) as our primary evaluation metrics.

\paragraph{Current retrievers capture information from limited perspectives.}
We examine the zero-shot performance of naive retrievers on \textsc{Multi$^3$IR}. They reach only up to 28.83 HC@10 and 39.52 SC@10, leaving a substantial gap to the oracle setting where the perspective descriptions are explicitly provided. Providing these perspectives boosts performance by 15.58 to 35.31 points on HC@10 and by 31.28 to 36.72 points on SC@10, suggesting that current retrievers struggle to identify perspectives that are only implicit in a query. The gap persists even at $k=100$, with a difference of 19.45 to 35.60 points on HC@100, indicating that the failure is not one of retrieval depth but of query understanding.


\paragraph{How the retriever learns matters more than what it learns from.}
We compare methods within each level of supervision. Under document-level supervision, \textsc{ARE} improves over the fine-tuned naive retriever by 5.30 to 6.20 points on HC@10. Under perspective-level supervision, the gap widens: \textsc{SPIN} outperforms \textsc{LLM-Expansion} by 8.18 to 13.13 points on HC@10 and by 17.99 to 21.92 points on SC@10, despite using the same annotations and the same number of query embeddings. Notably, multiple query embeddings alone provide no benefit, as \textsc{LLM-Expansion} falls below the naive baseline on HC@10 while also using $m=5$. Multi-vector retrieval helps only when the embeddings are explicitly trained to align with distinct perspectives, indicating that the architecture and training objective, rather than the supervision signal alone, drive the improvement.

\paragraph{Perspective-guided learning boosts retrieval with lighter annotation effort.}
We compare \textsc{ARE} and \textsc{SPIN}, two multi-vector retrievers trained with document-level and perspective-level supervision, respectively. As shown in Table~\ref{tab:main_results}, \textsc{SPIN} outperforms \textsc{ARE} by 3.62 to 4.47 points on HC@10 and by 14.47 to 14.62 points on SC@10, demonstrating that perspective-level alignment is more effective than document-level supervision for training multi-vector retrievers. Perspective-level supervision is also substantially cheaper to obtain, as document relevance annotation dominates the overall synthesis cost (Appendix~\ref{sec:annotation_cost}). Perspective-guided learning is thus both more effective and lighter to annotate, offering a more scalable path toward training perspective-aware retrievers.

\begin{table*}[t!]
    \centering
    \resizebox{\linewidth}{!}{%
    \begin{NiceTabular}{@{}>{\raggedright\arraybackslash}p{3cm}ccccccccccc@{}}[colortbl-like]
        \toprule
         & & & & \multicolumn{4}{c}{\textbf{PIR}} & \multicolumn{4}{c}{\textbf{BeRDS}} \\
         \cmidrule(r){5-8}\cmidrule(r){9-12}
        \textbf{Method} & $\mathbf{m}$ & \textbf{FT-Doc} & \textbf{FT-Persp} & @5 & @10 & @20 & @100 & @5 & @10 & @20 & @100 \\
        \midrule
        \textit{\textbf{MM-Embed}} & & & & & & & & & & & \\
        \text{Naive} & 1 & \xmark & \xmark & 51.12 & 59.98 & 69.29 & 87.41 & 79.10 & 88.85 & 94.95 & \textbf{99.15} \\
        \text{SPIN} & 5 & \xmark & \cmark & \textbf{63.33} & \textbf{70.25} & \textbf{80.55} & \textbf{93.50} & \textbf{83.45} & \textbf{90.20} & \textbf{96.80} & 98.10 \\
        \midrule
        \textit{\textbf{GME-Qwen2}} & & & & & & & & & & & \\
        \text{Naive} & 1 & \xmark & \xmark & 49.58 & 58.02 & 66.50 & 83.26 & 53.75 & 64.95 & 74.25 & 91.75 \\
        \text{SPIN} & 5 & \xmark & \cmark & \textbf{59.11} & \textbf{69.75} & \textbf{76.63} & \textbf{89.08} & \textbf{59.75} & \textbf{85.05} & \textbf{88.30} & \textbf{92.30} \\
        \midrule
        \textit{\textbf{Qwen3-VL}} & & & & & & & & & & & \\
        \text{Naive} & 1 & \xmark & \xmark & 51.06 & 60.10 & 69.96 & 87.16 & 75.05 & 84.30 & 89.90 & 96.15 \\
        \text{SPIN} & 5 & \xmark & \cmark & \textbf{55.21} & \textbf{69.47} & \textbf{78.36} & \textbf{90.97} & \textbf{76.20} & \textbf{86.90} & \textbf{92.40} & \textbf{96.80} \\
        \bottomrule
    \end{NiceTabular}}
    \caption{Perspective coverage on the unseen benchmarks, PIR~\cite{zhao2024beyond} and BeRDS~\cite{chen2025open}, measured by hard coverage (\%). SPIN is fine-tuned on the training split of Multi\textsuperscript{3}IR.}
    \label{tab:other_results}
\end{table*}
\subsection{Results on Other Benchmarks}
We report zero-shot retrieval performance on two unseen open-ended IR benchmarks, PIR~\citep{zhao2024beyond} and BeRDS~\citep{chen2025open}, in Table~\ref{tab:other_results}. Since both provide perspective-level gold documents, we measure HC@$k$ by strict document match, without a judge model. Each query is encoded from the original question alone, and the perspectives are never shown to the retriever. Further details on the benchmark subsets and corpora are in Appendix~\ref{sec:other_benchmark_details}.

\textsc{SPIN} improves over the naive retriever across all three backbones on both benchmarks. On PIR, the gains range from $9.37$ to $11.73$ points on HC@10 and remain substantial at $k{=}100$ ($3.81$ to $6.09$ points). On BeRDS, where the naive retrievers already achieve high coverage, the gains are smaller for \textsc{MM-Embed} and \textsc{Qwen3-VL} but reach $20.10$ points on HC@10 for \textsc{GME-Qwen2}, which has the lowest naive coverage. The only exception is \textsc{MM-Embed} at HC@100 on BeRDS, where both methods exceed 98\%, leaving little room for improvement. Overall, these results show that perspective-aware steering learned on \textsc{Multi$^3$IR} transfers to unseen benchmarks with different domains and corpora, without adaptation to the target distribution.

\section{Analysis}
All experiments in this section are conducted with Qwen3-VL-Embedding-8B~\citep{qwen3vlembedding}, which has 36 layers in total. Implementation details, including the training configuration and the retrieval instructions, follow Section~\ref{main_results}.

\begin{table}[t]
\centering
\small
\begin{NiceTabular}{lccccc}
    \toprule
     & \multicolumn{5}{c}{\textbf{Number of Perspectives ($m$)}} \\
     \cmidrule(r){2-6}
    \textbf{Layer} & 1 & 3 & 5 & 7 & 9 \\
    \midrule
    Early ($l=6$) & 56.7 & 58.5 & 58.1 & 56.8 & 57.2 \\
    Mid ($l=18$)  & 55.3 & 59.9 & 63.4 & 63.9 & 64.2 \\
    Late ($l=30$) & 62.1 & 62.8 & 61.0 & 60.5 & 60.7 \\
    \bottomrule
\end{NiceTabular}%
\caption{HC@100 score (\%) across the number of noise vectors ($m$) for different noise injection layers ($L$).}
\label{tab:layer_m}
\end{table}
\subsection{Effect of Injection Layer}
A central mechanism of \textsc{SPIN} is to inject noise into an intermediate layer of the retriever. As shown in Table~\ref{tab:layer_m}, injecting at the mid layer ($L{=}18$) yields consistent improvements as $m$ increases. In contrast, late-layer injection ($L{=}30$) plateaus as $m$ increases, likely because the representations at deeper layers have already collapsed toward a single direction. These results suggest that intermediate-layer injection is essential for \textsc{SPIN} to fully benefit from increasing $m$.

\begin{table}[t]
\centering
\small
\resizebox{\linewidth}{!}{%
\begin{tabular}{llrr}
\toprule
\textbf{Method} & \textbf{Layer} & \textbf{\# Params} & \textbf{HC@100} \\
\midrule
\multicolumn{4}{l}{\textbf{\textit{Perturbation-based}}} \\
SPIN ($m$=5)          & L0      & 20.5K  & 34.3 \\
SPIN ($m$=5)          & L18     & 20.5K  & \textbf{63.4} \\
Prefix-tuning ($k$=10)  & L0      & 204.8K & 48.1 \\
Prefix-tuning ($k$=10)  & L18     & 204.8K & 54.1 \\
\midrule
\multicolumn{4}{l}{\textbf{\textit{Reparameterization-based}}} \\
LoRA ($r$=16)         & L0--36  & 209.7M & 54.2 \\
LoRA ($r$=16)         & L18--36 & 104.9M & 55.9 \\
Adapter ($r$=64)      & L0--36  & 83.9M  & 53.9 \\
Adapter ($r$=64)      & L18--36 & 41.9M  & 55.9 \\
\bottomrule
\end{tabular}}
\caption{Perspective coverage (\%) comparison across steering methods and injection layers.}
\label{tab:steering_comparison}
\end{table}
\subsection{Comparison with Existing Adaptation Methods}
We isolate the methodological contribution of \textsc{SPIN} by comparing it with representative adaptation methods: perturbation-based methods such as prefix-tuning~\cite{li2021prefix}, and weight-based methods such as LoRA~\cite{hu2021lora} and adapter~\cite{houlsby2019parameter}.

\paragraph{Adaptation method.}
At the intermediate layer, \textsc{SPIN} achieves the best HC@100 ($63.4$) with only $20.5$K parameters, outperforming LoRA and adapter by $7.5$ points while using over three orders of magnitude fewer parameters. This indicates that learned additive vectors steer the query representation more effectively than reparameterizing the backbone weights.

\paragraph{Injection layer.}
We next examine how the injection layer affects each method. As shown in Table~\ref{tab:steering_comparison}, perturbation-based methods benefit substantially from intermediate-layer injection (\textsc{SPIN} $+29.1$, soft prompt $+6.0$), whereas reparameterization-based methods gain only marginally ($+1.7$ and $+2.0$) despite halving their parameters, as the frozen lower layers already provide strong representations.

\begin{table}[t]
\centering
\small
\begin{tabular}{lrrrr}
\toprule
\textbf{Method} & \textbf{@5} & \textbf{@10} & \textbf{@20} & \textbf{@100} \\
\midrule
\multicolumn{5}{l}{\textbf{\textit{Modality coverage}}} \\
Naive           & 26.9 & 32.4 & 38.9 & 48.9 \\
\textsc{SPIN}   & 40.2 & 53.0 & 59.7 & 78.2 \\
Oracle          & 50.5 & 56.3 & 64.6 & 81.1 \\
\midrule
\multicolumn{5}{l}{\textbf{\textit{Domain coverage}}} \\
Naive           & 18.7 & 27.3 & 37.7 & 53.4 \\
\textsc{SPIN}   & 30.6 & 42.7 & 50.2 & 68.3 \\
Oracle          & 46.0 & 61.7 & 73.1 & 84.8 \\
\bottomrule
\end{tabular}
\caption{Knowledge source coverage (\%). For each query, we measure the fraction of distinct modalities or domains in its gold documents that are covered by the top-$k$ retrieved results.}
\label{tab:coverage}
\end{table}
\subsection{Knowledge Source Coverage}
To quantify how well \textsc{SPIN} retrieves information across diverse knowledge sources spanning modalities and domains, we compare it with a zero-shot naive baseline and an oracle upper bound. As shown in Table~\ref{tab:coverage}, \textsc{SPIN} nearly saturates the oracle on modality coverage, reaching $78.2$ versus $81.1$ at $k{=}100$ and closing $91\%$ of the naive-to-oracle gap. In contrast, \textsc{SPIN} closes only $35$--$47\%$ of the naive-to-oracle gap for domain coverage, and the shortfall persists as $k$ grows ($68.3$ versus $84.8$ at $k{=}100$). Learning domain-aware steering vectors is a promising direction for closing this gap.

\begin{table}[t]
\centering
\small
\renewcommand{\arraystretch}{1.15}
\setlength{\tabcolsep}{3pt}
\resizebox{\linewidth}{!}{%
\begin{tabular}{l c c c c}
\toprule
& \multicolumn{4}{c}{\textbf{Cosine Distance Quartile}} \\
\cmidrule(lr){2-5}
\textbf{Method} & \textbf{$<$25\%} & \textbf{25--50\%} & \textbf{50--75\%} & \textbf{$>$75\%} \\
\midrule
Naive                        & 69.6 & 50.0 & 44.4 & 29.2 \\
SPIN, $m{=}1$                & 73.2 {\scriptsize(+3.6)} & 60.0 {\scriptsize(+10.0)} & 49.7 {\scriptsize(+5.3)} & 38.0 {\scriptsize(+8.8)} \\
SPIN, $m{=}3$                & 74.6 {\scriptsize(+5.0)} & 63.8 {\scriptsize(+13.8)} & 54.9 {\scriptsize(+10.5)} & 46.7 {\scriptsize(+17.5)} \\
SPIN, $m{=}5$                & 78.3 {\scriptsize(+8.7)} & 67.4 {\scriptsize(+17.4)} & 58.8 {\scriptsize(+14.4)} & 53.6 {\scriptsize(+24.4)} \\
\bottomrule
\end{tabular}%
}
\caption{HC@100 score (\%) across four quartiles of queries, grouped by how dispersed their positive documents are in the embedding space (average pairwise cosine distance). Parenthesized values denote gains over the Naive baseline.}
\label{tab:cos_dist}
\end{table}
\subsection{Effectiveness of \textsc{SPIN} on Clustered Positives}
Prior work identifies a limitation of dense retrievers: a single query vector may not be close to all relevant documents when they are scattered across the embedding space~\cite{chen2025beyond, weller2025theoretical}. To examine whether \textsc{SPIN} is effective for such dispersed queries, we analyze performance by document dispersion. As shown in Table~\ref{tab:cos_dist}, \textsc{SPIN} yields its largest gains on dispersed queries ($+24.4$ HC@100 at $m{=}5$ for the top quartile), while also improving substantially on tightly clustered queries ($+8.7$).

\section{Conclusion}
In this work, we introduce \textsc{Multi$^3$IR}, an open-ended IR benchmark of 104.9K queries from Stack Exchange, annotated with perspective descriptions and supporting documents. We further propose \textsc{SPIN}, a parameter- and label-efficient method that learns lightweight steering vectors to diversify a frozen retriever's output into multiple perspective-aware embeddings, using only perspective descriptions. Our experiments show that strong multimodal retrievers suffer from a single-perspective bias, collapsing onto one dominant viewpoint. \textsc{SPIN} substantially mitigates this bias, outperforms document-supervised baselines with lighter annotation, and generalizes to external benchmarks. Further analysis reveals that the bias persists even when relevant documents are tightly clustered in the embedding space, and that injecting steering vectors at an intermediate layer is critical to \textsc{SPIN}'s effectiveness.

\section*{Limitations}
We acknowledge several potential limitations of our work. (1) The number of perspectives can vary substantially across queries, while our method uses a fixed number of perspective vectors $m$. Developing an adaptive mechanism that selects $m$ based on each query is a promising direction for future work. (2) While \textsc{SPIN} reduces the need for document-level annotations, jointly leveraging perspective descriptions and their supporting documents during training could further improve perspective-specific retrieval. (3) Finally, \textsc{Multi$^3$IR} is constructed through an automatic pipeline involving perspective extraction, document retrieval, and exclusive-support verification. Although we validate this pipeline both quantitatively and qualitatively, LLM-induced biases may not be fully eliminated and could affect the absolute scores reported in our evaluations.

\section*{Acknowledgements}
We thank the anonymous reviewers for their valuable feedback.
This work was supported
by Institute of Information \& Communications Technology Planning \& Evaluation (IITP)
grants funded by the Korea government (MSIT)
(No.~RS-2025-02263841, Development of a Real-time Multimodal Framework for Comprehensive Deepfake Detection Incorporating Common Sense Error Analysis;
No.~RS-2019-II191082, SW StarLab;
No.~RS-2026-25524173, Ultra-Long-Term Hierarchical Memory and Reasoning Architecture for Next-Generation Omnimodal Agents;
No.~RS-2025-25442338, SNU AI Star Fellowship Support Program; 
and No.~RS-2021-II211343, SNU Artificial Intelligence Graduate School Program).
Gunhee Kim is the corresponding author. 

\bibliography{anthology,custom}
\bibliographystyle{acl_natbib}

\appendix
\newpage

\section{Subject Domain Classification}
\label{appendix:domain}
\begin{table*}[t!]
\centering
\small
\begin{tabularx}{\textwidth}{l>{\raggedright\arraybackslash}X}
\toprule
Top-level class & Sub-class \\
\midrule
Computer science & Computer science, knowledge \& systems; Bibliography; Library \& information sciences; Encyclopedias \& books of facts; Magazines, journals \& serials; Associations, organizations \& museums; News media, journalism \& publishing; Quotations; Manuscripts \& rare books \\
\addlinespace
Philosophy \& psychology & Philosophy; Metaphysics; Epistemology; Parapsychology \& occultism; Philosophical schools of thought; Psychology; Logic; Ethics; Ancient, medieval \& eastern philosophy; Modern western philosophy \\
\addlinespace
Religion & Religion; Philosophy \& theory of religion; The Bible; Christianity \& Christian theology; Christian practice \& observance; Christian pastoral practice \& religious orders; Christian organization, social work \& worship; History of Christianity; Christian denominations; Other religions \\
\addlinespace
Social sciences & Social sciences, sociology \& anthropology; Statistics; Political science; Economics; Law; Public administration \& military science; Social problems \& social services; Education; Commerce, communications \& transportation; Customs, etiquette \& folklore \\
\addlinespace
Language & Language; Linguistics; English \& Old English languages; German \& related languages; French \& related languages; Italian, Romanian \& related languages; Spanish \& Portuguese languages; Latin \& Italic languages; Classical \& modern Greek languages; Other languages \\
\addlinespace
Science & Science; Mathematics; Astronomy; Physics; Chemistry; Earth sciences \& geology; Fossils \& prehistoric life; Life sciences; biology; Plants (botany); Animals (zoology) \\
\addlinespace
Technology & Technology; Medicine \& health; Engineering; Agriculture; Home \& family management; Management \& public relations; Chemical engineering; Manufacturing; Manufacture for specific uses; Building \& construction \\
\addlinespace
Arts \& recreation & Arts; Landscaping \& area planning; Architecture; Sculpture, ceramics \& metalwork; Drawing \& decorative arts; Painting; Graphic arts; Photography \& computer art; Music; Sports, games \& entertainment \\
\addlinespace
Literature & Literature, rhetoric \& criticism; American literature in English; English \& Old English literatures; German \& related literatures; French \& related literatures; Italian, Romanian \& related literatures; Spanish \& Portuguese literatures; Latin \& Italic literatures; Classical \& modern Greek literatures; Other literatures \\
\addlinespace
History \& geography & History; Geography \& travel; Biography \& genealogy; History of the ancient world; History of Europe; History of Asia; History of Africa; History of North America; History of South America; History of other areas \\
\bottomrule
\end{tabularx}
\caption{Two-level subject domain taxonomy: 10 top-level classes, each divided into 10 sub-classes. Domains are labeled at the sub-class level.}
\label{tab:domains}
\end{table*}
We annotate each document's subject domain using the domain classifier\footnote{\url{https://huggingface.co/EssentialAI/eai-distill-0.5b}}, a text classification model that assigns a label from a two-level taxonomy of 10 top-level classes, each divided into 10 sub-classes, for 100 domains in total (Table~\ref{tab:domains}). Since the classifier accepts only text input, image documents require a textual surrogate. For WebQA, we use the caption paired with each image; for MultimodalQA, the title of the Wikipedia page from which the image is drawn; and for M-BEIR and MRMR, a caption generated by an image captioning model\footnote{\url{https://huggingface.co/Salesforce/blip-image-captioning-large}}. Since every document in $\text{Multi}^3\text{IR}$ originates from a web source, we crawl the source page for each image and use its textual content. Each input is truncated to the model's 512-token context window, and the domain with the highest predicted probability is assigned.

\section{Data Collection Details}

\subsection{Raw Data Collection}
\label{appendix:raw_data_collection}
\begin{table*}[t!]
\centering
\small
\begin{tabularx}{\textwidth}{l>{\raggedright\arraybackslash}Xrrrr}
\toprule
Category & Example Sites & \# Sites & Raw & After Quality Filter & After Sampling \\
\midrule
technology           & Stack Overflow, Super User, Unix \& Linux, Database Administrators    & 28 & 26{,}777{,}541 & 3{,}545{,}885 & 95{,}121 \\
culture \& recreation & English Language, Travel, Movies \& TV, Board Games     & 21 &     626{,}683 &     149{,}296 & 95{,}121 \\
science              & Mathematics, Physics, Chemistry, Biology          & 12 &   2{,}061{,}630 &     247{,}023 & 95{,}121 \\
business             & Personal Finance \& Money, Workplace, Project Management                  &  8 &     151{,}066 &      36{,}831 & 36{,}831 \\
life \& arts         & Home Improvement, Gardening, Fitness, Parenting                      &  8 &     168{,}113 &      31{,}707 & 31{,}707 \\
\midrule
Total                & ---                                               & 77 & 29{,}785{,}033 & 4{,}010{,}742 & 353{,}901 \\
\bottomrule
\end{tabularx}
\caption{Per-category counts of StackExchange questions at each stage of raw data collection: raw question count, after quality filtering (more than three answers and more than three distinct voters), and after category-level uniform sampling.}
\label{tab:raw_data_stats}
\end{table*}
We use the official StackExchange data dumps hosted on archive.org\footnote{\url{https://archive.org/details/stackexchange}}. Although the dataset is publicly available, it may contain user-generated content such as usernames, posts, and comments that could include personally identifying or offensive information. The authors checked all examples and found no personally identifying information (PII).

We select this source for three reasons: (1) the multi-answer structure provides natural multi-perspective supervision, (2) community voting and accept-tagging serve as built-in quality control, and (3) external links cited in answers serve as a natural source of perspective-grounded evidence.

We collect data from 77 sites organized into 5 broad categories: \textit{technology}, \textit{culture \& recreation}, \textit{science}, \textit{business}, and \textit{life \& arts}. Some sites appear in multiple categories under the multi-category specification (e.g., \texttt{graphic\_design} belongs to both \textit{technology} and \textit{life \& arts}). For each site, we retain a question only if it has more than three answers and receives votes from more than three distinct users, ensuring both quantity and diversity of viewpoints. For each retained post, the question $Q$ is formed by concatenating the title and body, and the answer set $A$ is the aggregation of all user responses. To avoid over-representation of dominant categories such as \textit{technology}, we then apply category-level uniform sampling with a per-category cap. Per-category counts at each stage with representative sites are reported in Table~\ref{tab:raw_data_stats}.

\subsection{Perspective Extraction and Verification}
\label{appendix:perspective_extraction_and_filtering}
\begin{figure*}[htbp]
\centering
\definecolor{inputcolor}{rgb}{0.10,0.35,0.70}
\definecolor{outputcolor}{rgb}{0.45,0.15,0.60}
\begin{tcolorbox}[
title=Perspective Extraction Prompt,
colback=gray!5,
colframe=gray!60,
width=0.95\textwidth
]

\small

Given a question and a list of answers, extract the distinct perspectives (facets) covered in the answers as independent sub-questions. Each sub-question should be:

\begin{enumerate}
    \item \textbf{Self-contained}: Understandable without the original question.
    \item \textbf{Search-friendly}: Suitable as a query to a search engine to find supporting evidence.
    \item \textbf{Specific}: Targeting exactly one perspective or facet from the answers.
    \item \textbf{Non-redundant}: Each sub-question should cover a distinct aspect.
\end{enumerate}

\medskip

\textbf{Output format} (one \texttt{<perspective>} block per distinct viewpoint)

\begin{tcblisting}{
colback=white,
colframe=gray!40,
listing only,
listing options={
    basicstyle=\ttfamily\small,
    breaklines=true,
    breakatwhitespace=true
}
}
<perspective>
{Search-friendly description of the perspective}
  - {Supporting fact or quote}
  - ...
</perspective>
\end{tcblisting}

\medskip

\textbf{Example input}

\begin{tcblisting}{
colback=white,
colframe=gray!40,
listing only,
listing options={
    basicstyle=\ttfamily\small\color{inputcolor},
    breaklines=true,
    breakatwhitespace=true
}
}
<question>
Why does my character walk forward automatically? ...
</question>

A1: Minecraft lost the "key W was released" event, so you're still going forwards. ...
A2: One cause is pressing a movement key + a GUI key simultaneously -- a "poor-man's autorun" that keeps you walking after closing the inventory. ...
A3: A known bug when moving and clicking at the same time. Workaround: update the LWJGL library. ...
\end{tcblisting}

\medskip

\textbf{Example output}

\begin{tcblisting}{
colback=white,
colframe=gray!40,
listing only,
listing options={
    basicstyle=\ttfamily\small\color{outputcolor},
    breaklines=true,
    breakatwhitespace=true
}
}
<perspective>
Could Minecraft lose a "key released" event so the game keeps thinking W is held down?
  - Minecraft lost the "key W was released" event.
</perspective>
<perspective>
Can pressing a movement key and a GUI key simultaneously cause persistent forward walking?
  - A "poor-man's autorun" that keeps you walking after closing the inventory.
</perspective>
<perspective>
Is there a known LWJGL-related bug, and is updating LWJGL a documented workaround?
  - Known bug; workaround: update the LWJGL library.
</perspective>
\end{tcblisting}

\end{tcolorbox}

\caption{Prompt template for perspective extraction. {\color{inputcolor}Blue text} denotes instance-specific input, and {\color{outputcolor}violet text} denotes the model's expected output.}
\label{fig:prompt_perspective_extraction}
\end{figure*}
\begin{table}[t!]
\centering
\small
\resizebox{\columnwidth}{!}{%
\begin{tabular}{lrrr}
\toprule
Category & Questions & Perspectives & Avg.\ per Q \\
\midrule
culture \& recreation &  35{,}039 & 179{,}539 & 5.12 \\
science               &  20{,}465 &  95{,}937 & 4.69 \\
business              &  19{,}165 &  99{,}408 & 5.19 \\
technology            &  16{,}085 &  74{,}213 & 4.61 \\
life \& arts          &  14{,}064 &  72{,}189 & 5.13 \\
professional          &      98 &     453 & 4.62 \\
\midrule
Total                 & 104{,}916 & 521{,}739 & 4.97 \\
\bottomrule
\end{tabular}%
}
\caption{Per-category distribution of retained questions and perspectives after extraction and filtering (train and evaluation splits combined).}
\label{tab:perspective_stats}
\end{table}

We extract perspectives using GPT-5-mini (gpt-5-mini-2025-08-07) via the OpenAI Batch API; the prompt and an example are shown in Figure~\ref{fig:prompt_perspective_extraction}. For uniqueness, we encode perspective summaries with the \texttt{sentence-transformers} library\footnote{\url{https://www.sbert.net/}} and iteratively remove a perspective whose cosine similarity to any previously kept perspective exceeds the threshold; the threshold of $0.8$ was chosen through manual inspection to best separate near-duplicates from genuinely distinct perspectives. For faithfulness, we concatenate the entire answer set into a single document and use it as the supporting document, and run the entailment check via the \texttt{minicheck} library\footnote{\url{https://github.com/Liyan06/MiniCheck}} on a vLLM backend\footnote{\url{https://github.com/vllm-project/vllm}} with prefix caching. The per-category distribution after both filters is reported in Table~\ref{tab:perspective_stats}.

\subsection{Document Retrieval and Verification}
\label{appendix:document_retrieval_and_filtering}
\begin{figure*}[t]
\centering
\begin{tcolorbox}[
title=Perspective--Document Supportedness Judge Prompt,
colback=gray!5,
colframe=gray!60,
width=0.95\textwidth
]

\small

You are judging whether a document supports answering a specific perspective.

Follow this reasoning process:

\begin{enumerate}
    \item Identify what the perspective specifically asks for.
    \item Describe what the document actually contains --- only what is literally visible. Do NOT assume or infer content.
    \item Compare the two and give a verdict.
\end{enumerate}

\textbf{Verdicts:}

\begin{itemize}
    \item \textbf{fully}: The document explicitly contains the specific answer the perspective asks for.
    \item \textbf{partially}: The document has related information but does not fully answer the specific question.
    \item \textbf{not}: The document is only topically related or unrelated, with no specific answer.
\end{itemize}

\textbf{Output constraints:}

\begin{itemize}
    \item Combine steps 1--3 into one paragraph inside \texttt{<reason>} tags.
    \item The verdict must be exactly one of: \texttt{fully}, \texttt{partially}, or \texttt{not}, inside \texttt{<verdict>} tags.
\end{itemize}

\medskip

\textbf{In-context examples}

\begin{tcblisting}{
colback=white,
colframe=gray!40,
listing only,
listing options={
    basicstyle=\ttfamily\small,
    breaklines=true,
    breakatwhitespace=true
}
}
Example 1.
<reason>The perspective asks for a method to spatially join point data to polygon attributes in QGIS. The document shows the QGIS "Join Attributes by Location" dialog [...]. This is the exact tool needed to perform the spatial join.</reason>
<verdict>fully</verdict>

Example 2. ...

Example 3. ...

Example 4. ...
\end{tcblisting}

\medskip

\textbf{Input format}

\begin{tcblisting}{
colback=white,
colframe=gray!40,
listing only,
listing options={
    basicstyle=\ttfamily\small,
    breaklines=true,
    breakatwhitespace=true
}
}
<document>{document}</document>
<perspective>{perspective}</perspective>
\end{tcblisting}

\medskip

\textbf{Output format}

\begin{tcblisting}{
colback=white,
colframe=gray!40,
listing only,
listing options={
    basicstyle=\ttfamily\small,
    breaklines=true,
    breakatwhitespace=true
}
}
<reason>your reasoning combining steps 1-3 in one paragraph</reason>
<verdict>fully|partially|not</verdict>
\end{tcblisting}

\end{tcolorbox}

\caption{Prompt and in-context examples used by the perspective--document supportedness judge. Examples 2--4 are abbreviated.}
\label{fig:prompt_support_judge}
\end{figure*}
\begin{table}[t]
\centering
\small
\resizebox{\columnwidth}{!}{%
\begin{tabular}{@{}llrrrr@{}}
\toprule
Model & Criterion & Agr. & Prec. & Rec. & Pass \\
\midrule
Qwen3-VL-Instruct & fully only        & 0.89 & 0.93 & 0.62 & 0.14 \\
                  & fully + partially & 0.77 & 0.78 & 0.79 & 0.35 \\
\midrule
Qwen3-VL-Thinking & fully only        & 0.89 & 0.95 & 0.48 & 0.10 \\
                  & fully + partially & 0.79 & 0.85 & 0.76 & 0.33 \\
\midrule
GPT-5-mini        & fully only        & 0.90 & 0.92 & 0.38 & 0.08 \\
                  & fully + partially & 0.54 & 0.70 & 0.93 & 0.71 \\
\bottomrule
\end{tabular}%
}
\caption{Document verifier performance across models and criteria, evaluated on 300 documents (150 text and 150 image) with human annotations as ground truth. Agr., Prec., and Rec. denote agreement, precision, and recall, respectively.}
\label{tab:judge_human}
\end{table}
\begin{table}[t]
\centering
\small
\begin{tabular}{@{}lrr@{}}
\toprule
Stage & \# Samples & Drop rate (\%) \\
\midrule
Document candidate          & 8,254,298 & -- \\
(1) Own-perspective support &  1,503,151 & 81.8 \\
(2) Sibling exclusivity     &  1,012,061 & 32.7 \\
\bottomrule
\end{tabular}
\caption{Document verification result with Qwen3-VL-Instruct. Drop rates are relative to the previous stage. Stage 1 requires \emph{full} support of the pair's own perspective; Stage 2 excludes documents also fully supporting a sibling perspective.}
\label{tab:document_verification_stat}
\end{table}

We use Qwen3-Embedding-4B\footnote{\url{https://huggingface.co/Qwen/Qwen3-Embedding-4B}} served via vLLM to encode both the query and the C4\footnote{\url{https://huggingface.co/datasets/allenai/c4}} English documents with a 4{,}096-token context window. Queries are prefixed with \texttt{``Instruct: Given a web search query, retrieve relevant passages that answer the query\textbackslash nQuery:\{query\}''}, while documents are encoded without any prefix. Image candidates are obtained from the Google Images endpoint of the Serper API\footnote{\url{https://serper.dev/}}.

To verify whether each document exclusively supports its assigned perspective, we prompt a VLM using the template in Figure~\ref{fig:prompt_support_judge}. We evaluate three models under two criteria: (i) \textit{fully-only}, which accepts only \textit{fully support} predictions, and (ii) \textit{fully+partially}, which also accepts \textit{partially support}. The former shows higher agreement with human judgments, while Qwen3-VL-30B-A3B-Instruct\footnote{\url{https://huggingface.co/Qwen/Qwen3-VL-30B-A3B-Instruct}} achieves the best precision and recall.

\subsection{Human Annotation Procedure}
\label{appendix:human_annotation}
\begin{figure*}[!t]
    \centering
    \includegraphics[width=1\linewidth]{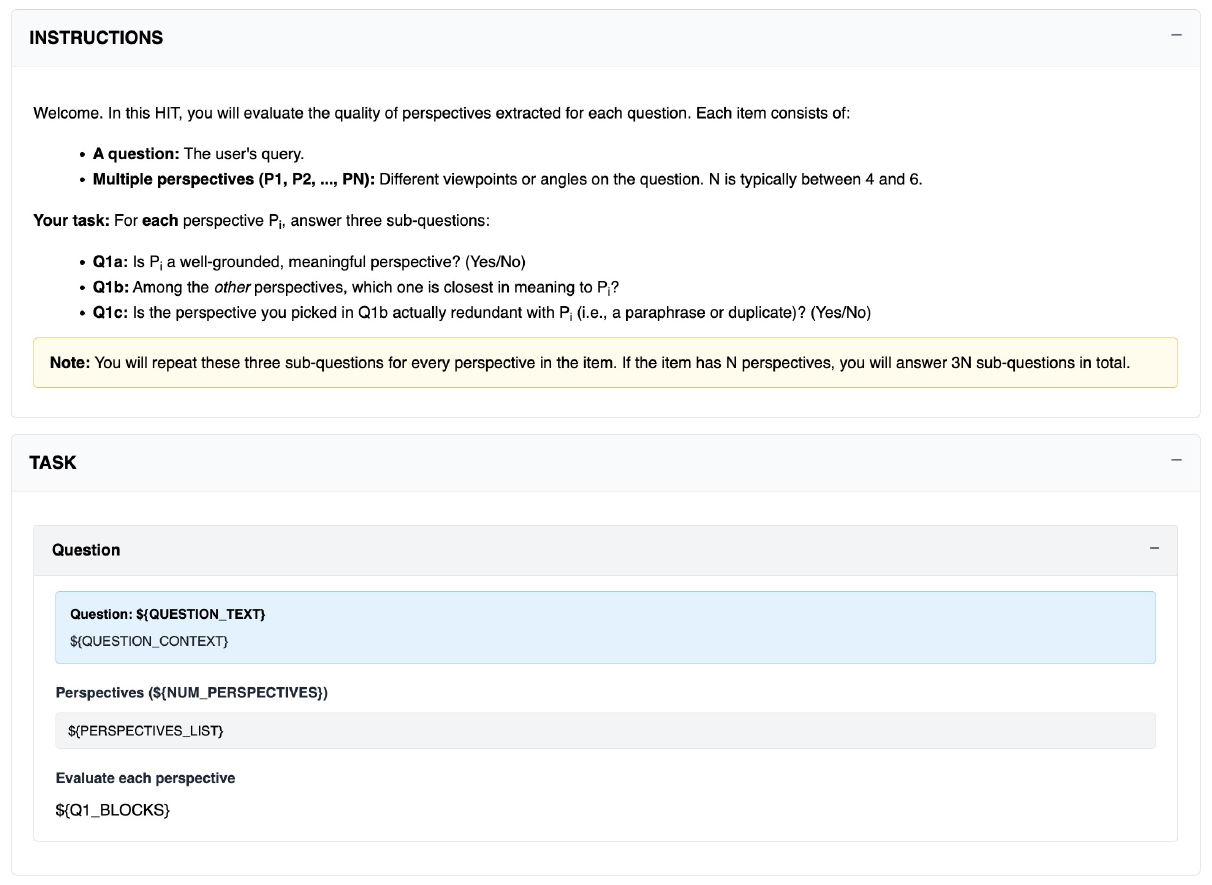}
    \caption{Instructions provided to Amazon Mechanical Turk raters for human annotation of perspective quality and redundancy.} \label{fig:annotation_perspective}
\end{figure*}
\begin{figure*}[!t]
    \centering
    \includegraphics[width=1\linewidth]{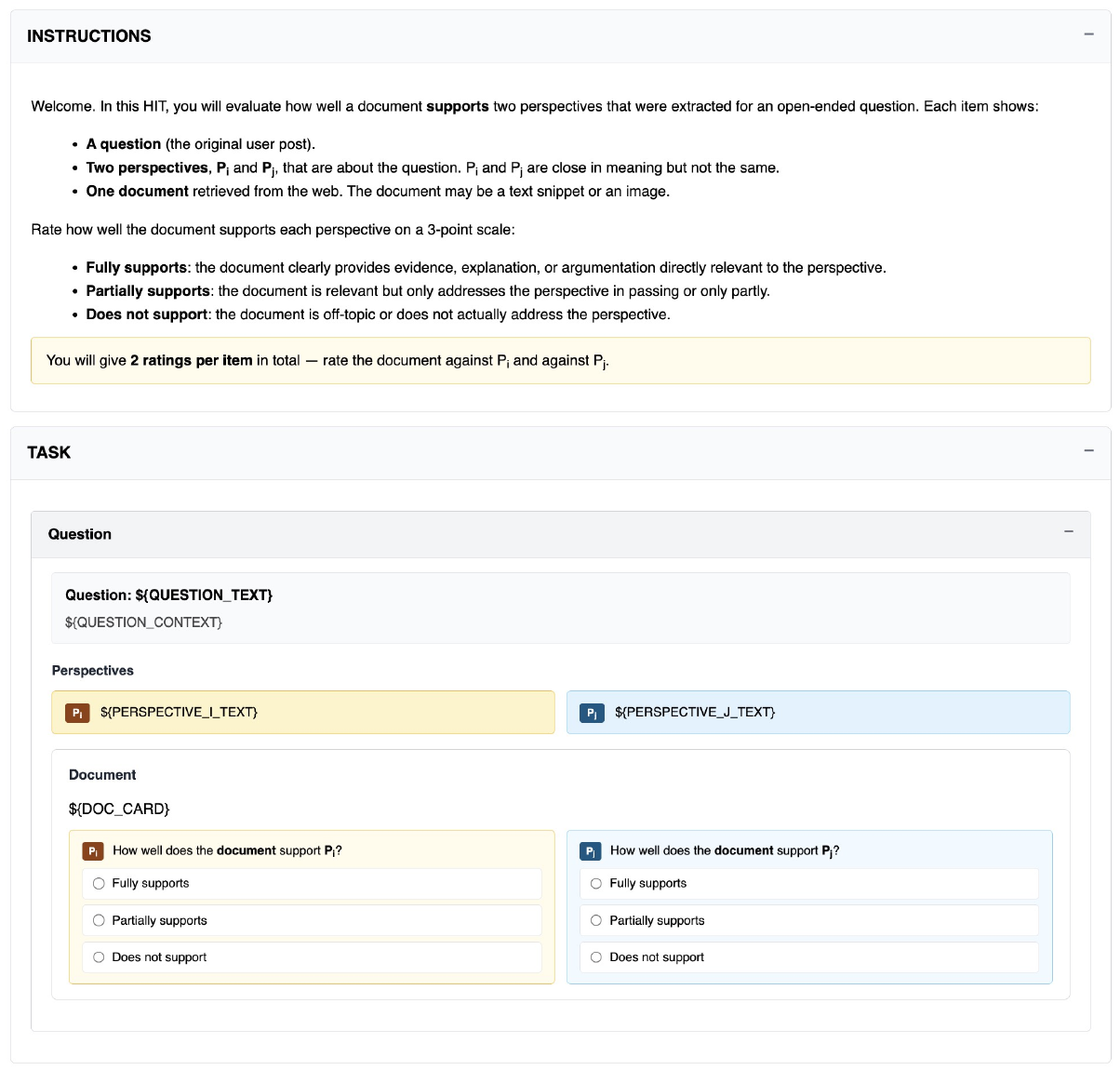}
    \caption{Instructions provided to Amazon Mechanical Turk raters for human annotation of perspective--document supportedness.} \label{fig:annotation_document}
\end{figure*}
\begin{table}[ht]
\centering
\small
\resizebox{\linewidth}{!}{%
\begin{tabular}{lccc}
\toprule
\text{Metric} & \text{N} & \text{Result} & \text{Fleiss' $\kappa$} \\
\midrule
Perspective relevance & 5.0K & 0.99 & 0.72 \\
Document exclusive-support & 9.1K & 0.90 & 0.59 / 0.72 \\
Answer coverage & 100 & 0.99 & 0.81 \\
Perspective faithfulness & 100 & 0.98 & 0.88 \\
\bottomrule
\end{tabular}%
}
\caption{Human verification of the automated pipeline. For document exclusive-support, Fleiss' $\kappa$ is reported as 3-class / binary.}
\label{tab:human_verification}
\end{table}
We hire skilled raters on Amazon Mechanical Turk (MTurk) to verify the quality of extracted perspectives and the exclusivity of their paired documents. The verification proceeds in two stages: a \emph{perspective evaluation} stage (Q1a/Q1b/Q1c) and a \emph{document--perspective support} stage.

Annotators were selected based on their success in qualification HITs that assessed their ability to judge whether a perspective is meaningful and whether two perspectives express the same underlying idea. This qualification task consisted of 3 verified examples with 5 sub-questions each (15 sub-questions in total), with a payment of \$2.00. We required annotators to be from English-speaking countries, have completed more than 10,000 HITS, and maintain a HIT approval rate greater than 95\%. 

After qualification, annotators received detailed instructions and proceeded to the perspective evaluation HIT. For each perspective $p_i$ in a question with $4 \le N \le 6$ extracted perspectives, they answered three sub-questions: (Q1a) whether $p_i$ is a meaningful perspective on the question (yes/no); (Q1b) which of the remaining $N{-}1$ perspectives is closest in meaning to $p_i$; and (Q1c) whether the perspective selected in Q1b is redundant with $p_i$, i.e., a paraphrase or duplicate (yes/no). Full annotator instructions appear in Figure~\ref{fig:annotation_perspective}. On an overlap subset of 30 questions per category, perspectives received unanimous \textit{yes} for Q1a (meaningful, 99.1\%) and \textit{no} for Q1c (unique, 96.2\%) across all domains. Q1b, an $(N{-}1)$-way classification harder than binary judgment, still reaches substantive agreement: $\kappa = 0.692$ (science) and $0.712$ (tech) for two-annotator domains, and $\kappa = 0.739$ (business), $0.740$ (culture), and $0.734$ (life \& arts) for three-annotator domains. The remaining 960 questions per category were annotated by a single annotator.

Next, we verify document--perspective support. For each retained perspective $p_i$, we identify its closest counterpart $p_j$ from Q1b, and each retrieved document originally paired with $p_i$ is judged on a 3-point scale (\textit{fully supports}, \textit{partially supports}, \textit{does not support}) against both $p_i$ and $p_j$. A document is retained only if it fully supports $p_i$ but does not fully support $p_j$, ensuring exclusivity; the perspective is discarded if no supporting document remains. On an overlap subset of 30 questions per category, Fleiss' $\kappa = 0.588$.

\subsection{Annotation Cost}
\label{sec:annotation_cost}
\begin{table}[t]
\centering
\small
\resizebox{\linewidth}{!}{%
\begin{tabular}{llrr}
\toprule
\text{Stage} & \text{Component} & \text{Compute (PFLOPs)} & \text{API Cost (\$)} \\
\midrule
\multirow{3}{*}{Persp.} & Extraction   & ---    & 0.238 \\
                             & Verification & 12.24  & ---   \\
                             & \textbf{Subtotal} & \textbf{12.24} & \textbf{0.238} \\
\midrule
\multirow{3}{*}{Doc.}    & Retrieval    & 104.64 & 0.364 \\
                             & Verification & 49.73  & ---   \\
                             & \textbf{Subtotal} & \textbf{154.37} & \textbf{0.364} \\
\bottomrule
\end{tabular}}
\caption{Stage-wise breakdown of supervision cost.}
\label{tab:cost_breakdown}
\end{table}
\begin{table}[t]
\centering
\small
\resizebox{\linewidth}{!}{%
\begin{tabular}{llrr}
\toprule
\text{Level} & \text{Included stage} & \text{Compute (PFLOPs)} & \text{API Cost (\$)} \\
\midrule
Persp.    & Persp.            & 12.24  & 0.238 \\
Doc. & Persp. + Doc. & 166.61 & 0.602 \\
\midrule
\textbf{Ratio} & & \textbf{13.6$\times$} & \textbf{2.5$\times$} \\
\bottomrule
\end{tabular}}
\caption{Overall cost by supervision level.}
\label{tab:cost_comparison}
\end{table}
We compare the annotation cost incurred at each stage of data synthesis. As shown in Tables~\ref{tab:cost_breakdown} and~\ref{tab:cost_comparison}, document relevance annotation dominates the overall cost, requiring 13.6$\times$ more local compute and 2.5$\times$ higher API cost than perspective-only supervision. Since SPIN requires only perspective-level supervision, it avoids this dominant cost component; our label-efficiency claim is therefore relative to methods that require document-level supervision.

\section{Experimental Details}

\subsection{Evaluation Setup}
\label{appendix:evaluation}
For the model-based judge, we use GPT-5-mini (gpt-5-mini-2025-08-07) via the OpenAI Chat Completions API\footnote{\url{https://platform.openai.com/docs/api-reference/chat}} with reasoning\_effort=medium and max\_completion\_tokens=2000. The judge is given a single perspective and one retrieved document at a time, and produces a chain-of-thought followed by a verdict of \textit{fully}, \textit{partially}, or \textit{not supported}; only \textit{fully} is counted as positive. The judge prompt is shown in Figure~\ref{fig:prompt_support_judge}.

To assess the reliability of the judge, we further measure its agreement with human annotators on a held-out sample of 100 perspective-document pairs, balanced across domains and modalities. We binarize the ratings into \textit{fully supported} and \textit{not supported}, where the latter includes both \textit{partially} and \textit{not supported}. Human annotators achieve substantial inter-annotator agreement (Fleiss' $\kappa=0.743$), while the GPT-5-mini judge shows even stronger agreement with the human majority vote (Fleiss' $\kappa=0.840$). These results indicate that the judge's decisions closely align with human judgments.

\subsection{Details on Other Benchmarks}
\label{sec:other_benchmark_details}

\paragraph{PIR.}
We use all six sub-datasets of PIR~\citep{zhao2024beyond}, comprising 4,187 questions and 10,019 perspectives. Each question is retrieved against its own corpus, and every perspective carries its own relevance judgments, so no subset selection is needed.

\paragraph{BeRDS.}
BeRDS~\citep{chen2025open} is designed to be scored by a judge model over retrieved passages rather than against gold documents, and only its ArguAna subset provides perspective-level document annotations. The Kialo and OpinionQA subsets define perspectives but leave them unannotated, so we restrict our evaluation to ArguAna (1,000 questions, 2,000 perspectives). Because the ArguAna document pool contains only 1,891 unique documents, we add the 355,123-passage Wikipedia corpus released with BeRDS as distractors.

\subsection{Implementation Details}
\label{appendix:implementation}

\subsubsection{Base retriever.} All methods for fine-tuning retrievers share the same instruction prompt, ``Retrieve images or text relevant to the user's query.'' All resulting embeddings are L2-normalized. Document embeddings are precomputed once and remain frozen. The main results are reported from a single run.

For GME-Qwen2-VL\footnote{\url{https://huggingface.co/Alibaba-NLP/gme-Qwen2-VL-7B-Instruct}} and Qwen3-VL-Embedding\footnote{\url{https://huggingface.co/Qwen/Qwen3-Embedding-8B}}, the instruction and the query are wrapped with the Qwen2-VL chat template\footnote{\url{https://huggingface.co/Qwen/Qwen2-VL-7B-Instruct/blob/main/chat_template.json}}, producing
\begin{quote}\small\ttfamily
\textless|im\_start|\textgreater system\textbackslash n\{instruction\}\textless|im\_end|\textgreater\\
\textless|im\_start|\textgreater user\textbackslash n\{query\}\textless|im\_end|\textgreater\\
\textless|im\_start|\textgreater assistant\textbackslash n\textless|endoftext|\textgreater
\end{quote}
Both are causal decoders, and we take the hidden state at the last token position as the query embedding.

For MM-Embed\footnote{\url{https://huggingface.co/nvidia/MM-Embed}}, which uses bidirectional attention, we instead concatenate the instruction and the query as \texttt{``Instruct: \{instruction\}\textbackslash nQuery: \{query\}''} following its original convention, and obtain the query embedding via the model's latent-attention pooling head, which aggregates the full sequence of hidden states through a learned attention layer.

\subsubsection{Fine-tuning}
All fine-tuned methods are trained on the same 72K queries sampled from \textsc{Multi$^3$IR} with AdamW~\citep{loshchilov2017decoupled} using a cosine schedule with $3\%$ warmup. Method-specific details are described below.

\paragraph{Naive (fine-tuned).} We fully fine-tune the retriever using InfoNCE with in-batch negatives (temperature $\tau{=}0.05$), with a learning rate of $1\mathrm{e}{-}5$ and a batch size of $32$ for $3$ epochs. The maximum query length is set to $512$ tokens.

\paragraph{LLM-Expansion.} We fully fine-tune Qwen3-4B\footnote{\url{https://huggingface.co/Qwen/Qwen3-4B}} on (query, perspective description) pairs with a learning rate of $2\mathrm{e}{-}5$ for $2$ epochs using a cosine schedule with $3\%$ warmup. The effective batch size is $16$ (batch size $4$ with $4$ gradient accumulation steps). At inference, we generate $m{=}5$ query reformulations per query using HuggingFace \texttt{transformers}\footnote{\url{https://github.com/huggingface/transformers}} with \texttt{AutoModelForCausalLM.generate}, with \textit{temperature}$=0.7$, \textit{top}-$p=0.8$, \textit{top}-$k=20$, and a maximum of $384$ tokens.

\paragraph{ARE.} We fine-tune the retriever with LoRA ($r{=}16$) and an additional input projection layer $\mathbf{W}_{\mathrm{in}} \in \mathbb{R}^{H \times H}$ ($H{=}3584$ for GME-Qwen2-VL, $H{=}4096$ for Qwen3-VL-Embedding; Xavier-initialized, without bias) that maps document embeddings into the language-model input space, using a learning rate of $3\mathrm{e}{-}5$, an effective batch size of $32$. Predicted query embeddings are aligned to gold document embeddings via Hungarian matching, and the matched pairs are trained with InfoNCE over in-batch negatives with temperature $\tau{=}0.01$ for Qwen3-VL-Embedding and $\tau{=}0.03$ for GME-Qwen2-VL.

\paragraph{\textsc{SPIN}.} We optimize $m$ noise vectors $\boldsymbol{\epsilon} \in \mathbb{R}^{m \times H}$ injected into the mid layer (layer $18$ of $36$ for MM-Embed and Qwen3-VL-Embedding, layer $14$ of $28$ for GME-Qwen2-VL), with all encoder weights frozen. We set $m{=}5$, yielding $20{,}480$ trainable parameters for MM-Embed and Qwen3-VL-Embedding ($H{=}4096$) and $17{,}920$ for GME-Qwen2-VL ($H{=}3584$). The noise vectors are initialized from $\mathcal{N}(0, 0.01^2)$. Learning rates are tuned per backbone: $3\mathrm{e}{-}4$ for MM-Embed, $1\mathrm{e}{-}2$ for Qwen3-VL-Embedding, and $1\mathrm{e}{-}3$ for GME-Qwen2-VL. We use a batch size of $32$. The training objective is the noisy-OR loss described in \S~\ref{perspective_guided}, where each similarity is converted into a probability via $p(d) = \sigma((\mathrm{sim} - b) / \tau)$ with $\tau{=}0.1$ and $b{=}0.5$ fixed across all backbones.

\subsubsection{Computational Resources}
All experiments are run on NVIDIA A6000 48GB GPUs (Intel Xeon Silver 4210R CPU). Naive (fine-tuned) is trained with $4$ GPUs using PyTorch Fully Sharded Data Parallel (FSDP)\footnote{\url{https://pytorch.org/docs/stable/fsdp.html}}, and LLM-Expansion is trained with $4$ GPUs using the TRL SFTTrainer\footnote{\url{https://huggingface.co/docs/trl/sft_trainer}} (which wraps HuggingFace Trainer's Distributed Data Parallel). ARE, \textsc{SPIN}, and zero-shot evaluation are run on a single GPU.

\section{Usage of GenAI}
AI-assisted coding tools were used to support data analysis and visualization, including drafting and debugging scripts for figure/table generation. All code and results were subsequently reviewed and validated by the authors.

\end{document}